\documentclass[showpacs,preprintnumbers,amsmath,amssymb,prl]{revtex4}
\usepackage{graphicx}
\usepackage{multirow}
\usepackage{dcolumn}
\usepackage{bm}
\usepackage{amsmath}
\usepackage{amsfonts}
\usepackage{amssymb}
\usepackage[hang]{subfigure}
\usepackage{multirow}
\usepackage{color}
\usepackage[bookmarks]{hyperref}

\newcommand{\ket}[1]{\mbox{$\mid \! #1 \, \rangle$}}
\newcommand{\bra}[1]{\mbox{$\langle \, #1 \! \mid$}}

\newcommand{\ketbra}[1]{\!\mid \! #1 \, \rangle\langle \, #1 \! \mid}

\newcommand{\eea}{\end{eqnarray}}
\newcommand{\bea}{\begin{eqnarray}}
\newcommand{\eeas}{\end{eqnarray*}}
\newcommand{\beas}{\begin{eqnarray*}}

\newcommand{\sx}{\sigma_x}
\newcommand{\sy}{\sigma_y}
\newcommand{\sz}{\sigma_z}

\newcommand{\ii}{\leavevmode\hbox{\.{\kern-.28em\tiny l\tiny\kern-.18em l\phantom{.}}}}

\begin{document}

\title{Experimental implementation of a four-player quantum game}

\author{Christian Schmid$^{1,2}$}
\author{Adrian P. Flitney$^{3}$}
\author{Witlef Wieczorek$^{1,2}$}
\author{Nikolai Kiesel$^{1,2}$}
\author{Harald Weinfurter$^{1,2}$}
\author{Lloyd C. L. Hollenberg$^{3,4}$}

\affiliation{$^{1}$Sektion Physik,
Ludwig-Maximilians-Universit{\"a}t, D-80797 M{\"u}nchen, Germany}
\affiliation{$^{2}$Max-Planck-Institut f{\"u}r Quantenoptik, D-85748
Garching, Germany} \affiliation{$^{3}$School of Physics, The
University of Melbourne, Melbourne, Victoria 3010, Australia}
\affiliation{$^{4}$Centre for Quantum Computing, University of
Melbourne, Melbourne, Victoria 3010, Australia}

\date{\today}

\begin{abstract}
Game theory is central to the understanding of competitive
interactions arising in many fields, from the social and physical
sciences to economics. Recently, as the definition of information is
generalized to include entangled quantum systems, quantum game
theory has emerged as a framework for understanding the competitive
flow of quantum information. Up till now only two-player quantum
games have been demonstrated. Here we report the first experiment
that implements a four-player quantum Minority game over tunable
four-partite entangled states encoded in the polarization of single
photons. Experimental application of appropriate quantum player
strategies give equilibrium payoff values well above those
achievable in the classical game. These results are in excellent
quantitative agreement with our theoretical analysis of the
symmetric Pareto optimal strategies. Our result demonstrate for the
first time how non-trivial equilibria can arise in a competitive
situation involving quantum agents and pave the way for a range of
quantum transaction applications.
\end{abstract}


\maketitle

\section{Introduction}
Originally developed for economics, the impact of game theory in
this field is now pervasive, and has spread to many other
disciplines as diverse as evolutionary biology~\cite{May82}, the
social sciences~\cite{Bed07} and international
conflict~\cite{Pow02}. Game theory is characterized by the solution
concepts of the \emph{ Nash equilibrium} (NE), the result from which
no player can improve their payoff by a unilateral change in
strategy, and the \emph{Pareto optimal} (PO) outcome, the one from
which no player can improve their payoff without another player
being worse off. The former can be considered as the equilibrium
that is best for each player individually, while the latter is
generally the best result for the players as a group.

In the search for new applications of quantum information processing
based on competitive interactions between agents the appropriate
language is quantum game theory. Applications can be based on
exchanges between several agents who share entanglement, in contrast
to direct classical communication. The sharing of an entangled
resource permits competitive von Neumann-type games, with
applications such as quantum auctions~\cite{Hog07}, quantum
voting~\cite{Vac07}, and quantum communication. This is distinct
from cooperative games that arise where agents are permitted to
communicate, either directly or through a third party. In quantum
games, entanglement is used as a resource that can enhance the
payoffs of well known game-theoretic equilibria~\cite{Ben01a}.

While there have been implementations of two-player quantum games
with limited strategic spaces~\cite{Du02,Pre07}, here we report for the
first time the implementation of a genuine multiplayer quantum game
driven by a four particle entangled state. Generally, two-player
quantum games do not see the enhancement of the pure strategy
equilibrium payoffs~\cite{Eis00,Ben01b} that can arise in
multiplayer quantum games. The practical exploration of quantum
games allows us to learn about the nature of quantum information in
a real situation. Figure~\ref{fig:game} compares the classical and
quantum versions of a four-player game.

\setlength{\unitlength}{0.5pt}
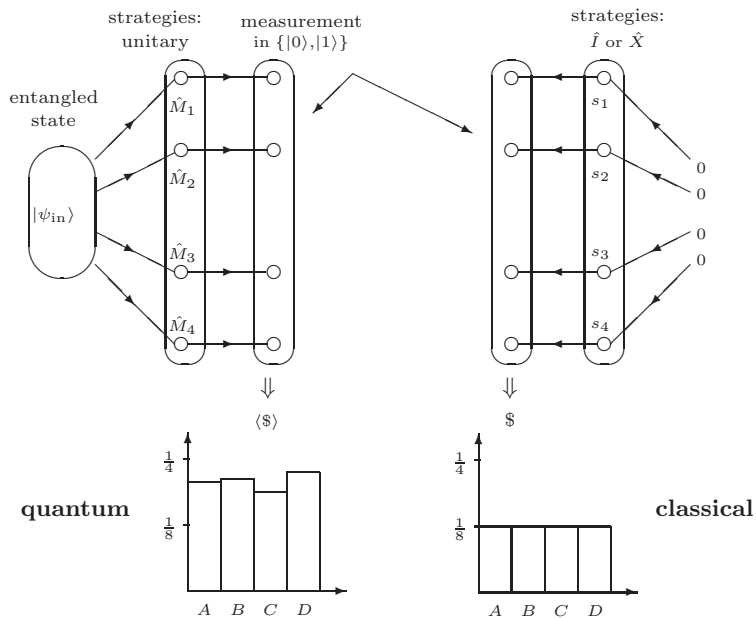
\begin{figure}
\begin{center}
\begin{picture}(565,400)(0,-20)

\put(0,120){
%
%
	\begin{picture}(215,300)(0,-60)
	\put(25,100){\oval(50,100)}
	\put(50,60){\vector(1,-1){30}}
	\put(80,30){\line(1,-1){30}}
	\put(50,85){\vector(2,-1){30}}
	\put(80,70){\line(2,-1){30}}
	\put(50,115){\vector(2,1){30}}
	\put(80,130){\line(2,1){30}}
	\put(50,140){\vector(1,1){30}}
	\put(80,170){\line(1,1){30}}
	\multiput(115,0)(0,55){2}{\circle{10}}
	\multiput(115,147)(0,55){2}{\circle{10}}
	\put(105,176){$\scriptstyle \hat{M}_1$}
	\put(105,121){$\scriptstyle \hat{M}_2$}
	\put(105,65){$\scriptstyle \hat{M}_3$}
	\put(105,10){$\scriptstyle \hat{M}_4$}

	\put(120,202){\vector(1,0){35}}
	\put(155,202){\line(1,0){25}}
	\put(120,147){\vector(1,0){35}}
	\put(155,147){\line(1,0){25}}
	\put(120,55){\vector(1,0){35}}
	\put(154,55){\line(1,0){25}}
	\put(120,0){\vector(1,0){35}}
	\put(155,0){\line(1,0){25}}

	\multiput(185,0)(0,55){2}{\circle{10}}
	\multiput(185,147)(0,55){2}{\circle{10}}

	\put(2,95){$\scriptstyle |\psi_\text{in} \rangle$}

	\put(118,100){\oval(30,230)}
	\put(185,100){\oval(30,230)}

	\put(-15,165){\shortstack{\scriptsize entangled\\ \scriptsize state}}
	\put(60,225){\shortstack{\scriptsize strategies:\\ \scriptsize unitary}}
	\put(175,-38){$\Downarrow$}
	\put(170,-62){$\scriptstyle \langle \$ \rangle$}
	\put(160,225){\shortstack{\scriptsize measurement \\ $\scriptstyle \text{in} \; \{|0\rangle, |1\rangle \}$}}

	\end{picture}}

\put(365,120){
%
%
	\begin{picture}(200,300)(15,-60)
	\put(150,60){\vector(-1,-1){30}}
	\put(120,30){\line(-1,-1){30}}
	\put(150,85){\vector(-2,-1){30}}
	\put(120,70){\line(-2,-1){30}}
	\put(150,115){\vector(-2,1){30}}
	\put(120,130){\line(-2,1){30}}
	\put(150,140){\vector(-1,1){30}}
	\put(120,170){\line(-1,1){30}}
	\multiput(85,0)(0,55){2}{\circle{10}}
	\multiput(85,147)(0,55){2}{\circle{10}}
	\put(75,181){$\scriptstyle s_1$}
	\put(75,126){$\scriptstyle s_2$}
	\put(75,67){$\scriptstyle s_3$}
	\put(75,12){$\scriptstyle s_4$}

	\multiput(155,60)(0,20){2}{$\scriptstyle 0$}
	\multiput(155,110)(0,20){2}{$\scriptstyle 0$}

	\put(80,202){\vector(-1,0){35}}
	\put(45,202){\line(-1,0){25}}
	\put(80,147){\vector(-1,0){35}}
	\put(45,147){\line(-1,0){25}}
	\put(80,55){\vector(-1,0){35}}
	\put(45,55){\line(-1,0){25}}
	\put(80,0){\vector(-1,0){35}}
	\put(45,0){\line(-1,0){25}}

	\multiput(15,0)(0,55){2}{\circle{10}}
	\multiput(15,147)(0,55){2}{\circle{10}}

	\put(85,100){\oval(30,230)}
	\put(15,100){\oval(30,230)}
	\put(8,-38){$\Downarrow$}
	\put(10,-62){\scriptsize \$}
	\put(60,225){\shortstack{\scriptsize strategies:\\ $\scriptstyle \hat{I} \; \text{or} \; \hat{X}$}}

	\end{picture}}

\put(100,-25){
%
%
	\begin{picture}(120,120)(-10,-8)

	\put(10,10){\vector(1,0){120}}
	\put(10,10){\vector(0,1){120}}
	\put(17,-8){$\scriptstyle A$}
	\put(42,-8){$\scriptstyle B$}
	\put(67,-8){$\scriptstyle C$}
	\put(92,-8){$\scriptstyle D$}
	\multiput(7,60)(0,50){2}{\line(1,0){4}}
	\put(-10,55){$\scriptstyle \frac{1}{8}$}
	\put(-10,105){$\scriptstyle \frac{1}{4}$}
	\put(35,10){\line(0,1){85}}
	\put(60,10){\line(0,1){85}}
	\put(85,10){\line(0,1){90}}
	\put(110,10){\line(0,1){90}}
	\put(10,93){\line(1,0){25}}
	\put(35,95){\line(1,0){25}}
	\put(60,85){\line(1,0){25}}
	\put(85,100){\line(1,0){25}}

	\end{picture}}

\put(320,-25){
%
%
	\begin{picture}(140,120)(-10,-7)

	\put(10,10){\vector(1,0){120}}
	\put(10,10){\vector(0,1){120}}
	\put(17,-8){$\scriptstyle A$}
	\put(42,-8){$\scriptstyle B$}
	\put(67,-8){$\scriptstyle C$}
	\put(92,-8){$\scriptstyle D$}
	\multiput(7,60)(0,50){2}{\line(1,0){4}}
	\put(-10,55){$\scriptstyle \frac{1}{8}$}
	\put(-10,105){$\scriptstyle \frac{1}{4}$}
	\put(35,10){\line(0,1){50}}
	\put(60,10){\line(0,1){50}}
	\put(85,10){\line(0,1){50}}
	\put(110,10){\line(0,1){50}}
	\put(10,60){\line(1,0){25}}
	\put(35,60){\line(1,0){25}}
	\put(60,60){\line(1,0){25}}
	\put(85,60){\line(1,0){25}}


	\end{picture}}

%
%
\put(245,355){
	\begin{picture}(100,30)(0,0)
	\put(0,30){\vector(2,-1){90}}
	\put(0,30){\vector(-1,-1){30}}
	\end{picture}
}
\put(0,50){\bf quantum}
\put(480,50){\bf classical}
\end{picture}

\end{center}
\caption{Schematic showing the flow
of information in a four-player quantum game (left) and a
four-player classical game (right). In the quantum version the players
utilize an entangled state as a resource, however in both versions
there is no communication between the players. For the particular
case of the Minority game the equilibrium payoffs are indicated in
the bar graph, with those of the quantum case being the values
achieved experimentally with our setup.}
\label{fig:game}
\end{figure}

\begin{figure}
\includegraphics[width=8.5cm,keepaspectratio]{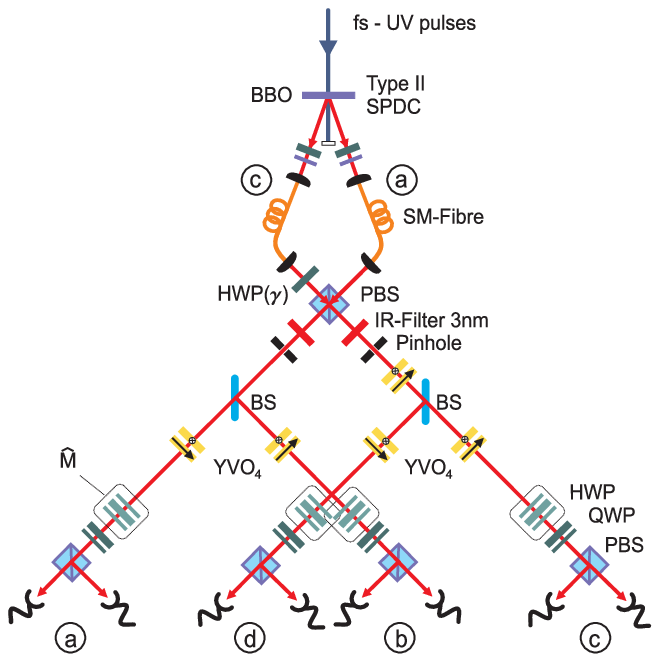}
\caption{Setup for the observation of $\ket{\Psi(\alpha)}$. Four
photons, emitted in two spatial modes $a$ and $c$, interfere at a
polarizing beam splitter (PBS). Prior to the interference the
polarization of the photons in mode $c$ is rotated by a
half-waveplate (HWP) set to an angle $\gamma$. Afterwards the
photons are split into four output modes $a,\,b,\,c,\,d$ by two beam
splitters (BS). Under the condition of detecting one photon in each
mode the desired state is observed.
The player unitary strategies are executed by a combination of a quarter-, a half- 
and a second quarter-waveplate ($\hat{M}$).
}
\label{fig:setup}
\end{figure}

We choose to implement a four-player Minority game.
The Minority game arises as a simple multi-agent model for
studying strategic decision making within a group of agents~\cite{Cha97}.
Its classical version has been used as an iterative model of buying and selling in a
stockmarket~\cite{Joh98,Sav99,Cha00}.
In each play, the agents independently select between one of two options, labeled
`0' and `1' (`buy' and `sell').
Those that choose the minority are awarded a payoff of one unit,
while the others receive no payoff.
Players can utilize knowledge of past successful choices to
optimize their strategy.
In the quantum situation we can represent each player's binary choice by the polarization of a photon
as in Fig.~\ref{fig:setup}.
Quantum versions of a one-shot Minority
game have generated some theoretical interest~\cite{Ben01a,Che04,Fli07b,Fli07a,Fli07d}.
In this paper we report on the first experimental implementation of the quantum Minority game (QMG).

We begin with a theoretical exploration of the QMG.
With the particular experiment in mind,
we choose the initial state to belong to a continuous set of four-partite
entangled states involving a mixture of the GHZ state and products
of EPR pairs.
We calculate the NE and PO strategy profiles as a function of the initial state.
Our method is a general one that can be applied to any quantum game
that has a similar protocol.
Our quantization of the Minority game is described as follows
and is shown schematically in Fig.~\ref{fig:game}.
Each of four players is given one qubit from a known four-partite
entangled state. This state is an element of the subspace spanned
by the four-qubit GHZ state and a product state of two Bell (or
EPR) pairs. The players are permitted to act on their qubit with
any local unitary operation. The choice of such an operator is the
player's strategy. During this stage coherence is maintained and
no communication between the players is permitted. After the
player actions, a referee measures the qubits in the computational
basis and awards payoffs using the classical payoff scheme. In the
dominant protocol of quantum games due to Eisert \emph{et al.}~\cite{Ben01a,Eis99a},
an entangling gate is used to produce a
GHZ state from an initial state of $\ket{00 \dots 0}$. After the
players' actions the inverse operator is applied to the
multi-partite state. For the Minority game this last operator only
has the effect of interchanging between states where the same
player(s) win and so can be omitted without changing the expected payoffs.
Our arrangement is consistent with the generalized quantum game formalisms of Lee and
Johnson~\cite{Lee03} and Gutoski and Watrous~\cite{Gut07}, and is
particularly suited for an implementation using entangled photon
states and linear optics quantum logic.


\section{Theory}\label{sec:theory}

A four player QMG was first examined by Benjamin and
Hayden~\cite{Ben01a}, and later generalized to multiple
players~\cite{Che04} and to the consideration of
decoherence~\cite{Fli07b}. Formally the QMG is played by computing
the state
\begin{equation}
\label{eqn:final} \rho_\text{final} = (\hat{A} \otimes \hat{B}
\otimes \hat{C} \otimes \hat{D}) \rho_\text{in}
        (\hat{A} \otimes \hat{B} \otimes \hat{C} \otimes \hat{D})^{\dagger},
\end{equation}
where $\hat{A}, \hat{B}, \hat{C}, \hat{D}$ are the operators
representing the strategies of the four players and $
\rho_\text{in}$ is some four qubit input state. The qubits in
$\rho_\text{final}$ are subsequently measured in the computation
basis and payoffs are awarded using the classical payoff matrix.
Work to date has concentrated on the consideration of an initial
GHZ state. In this paper we instead consider an the initial state
that is a superposition of the GHZ state with products of EPR
pairs:
\begin{equation}
\label{eqn:GHZ_EPR}
\begin{split}
        \ket{\Psi(\alpha)}&=\frac{\alpha}{\sqrt{2}}(\ket{0000} \:+\: \ket{1111})
        + \frac{\sqrt{1 - \alpha^2}}{2} \, (\ket{01} + \ket{10}) \otimes (\ket{01} + \ket{10}) \\
      &\equiv \frac{\alpha}{\sqrt{2}}\left(\ket{HHHH}+\ket{VVVV}
        \right) \\
        & + \frac{\sqrt{1-\alpha^2}}{2} \Bigl(\ket{HVHV}+\ket{HVVH}
        + \ket{VHHV}+\ket{VHVH},
        \Bigr)
\end{split}
\end{equation}
where $\alpha \in [0,1]$ is an adjustable parameter.
The qubits are encoded in the polarization of
single photons propagating in well-defined spatial modes.
The computational basis states $\ket{0}$ and $\ket{1}$ are represented
by the states $\ket{H}_i$ and $\ket{V}_i$, denoting the state of a
single photon in the spatial mode $i$ with linear horizontal and
linear vertical polarization, respectively. Most of the time the
spatial mode will be evident from the context and hence the
subscript be omitted.

To make allowances for some loss of fidelity in the preparation of this
state the initial state will be written as the mixed state
\begin{equation}
\label{eqn:noise}
\rho_\text{in} = f \ket{\Psi} \bra{\Psi}
    + \frac{1-f}{16} \Sigma_{ijk\ell = 0,1} \, \ket{ijk\ell} \bra{ijk\ell},
\end{equation}
where $f \in [0,1]$ is a measure of the fidelity of production of
the desired initial state. The second term in
\eqref{eqn:noise} represents completely random noise~\cite{Wei01}.
The players' strategies are single-qubit unitary operators that
can be parametrized in the form
\begin{equation}
\label{e-su2} \hat{M}(\theta, \beta_1, \beta_2) =
    \left( \begin{array}{cc} e^{i \beta_1} \cos(\theta/2) &   i e^{i \beta_2} \sin(\theta/2) \\
                     i e^{-i \beta_2} \sin(\theta/2) & e^{-i \beta_1} \cos(\theta/2)
    \end{array} \right),
\end{equation}
where $\theta \in [0, \pi], \, \beta_1, \beta_2 \in [-\pi, \pi]$.
In our construction of the QMG only the difference in the phases
is relevant to the expected payoff so it suffices to use a
restricted set of operators where $\beta \equiv \beta_1 =
-\beta_2$.

In the case where $\alpha=1$ it is known that a symmetric NE
occurs when all players choose the
strategy~\cite{Ben01a}
\begin{equation}
\label{eqn:GHZ_NE} \hat{M}(\frac{\pi}{2}, \frac{\pi}{8},\frac{-\pi}{8}) =
    \frac{1}{\sqrt{2}} \left( \begin{array}{cc} e^{i \pi/8} & i e^{-i \pi/8} \\
                                                              i e^{i \pi/8} & e^{-i \pi/8}
                                      \end{array} \right).
\end{equation}
Although the value $\beta_1 = -\beta_2 = \pi/8$ is not the unique
optimum, it is a focal point~\cite{Sch60} that attracts the
attention of the players since it is the simplest optimum value,
and therefore there is no great difficulty in arriving at this NE.
Given that at most one player of the four can be in the minority,
$\frac{1}{4}$ is the greatest average payoff that can be expected.
This is realized with the above strategy when the initial state
has maximum fidelity. As $f \rightarrow 0$, the payoff reduces to
$\frac{1}{8}$, the optimal payoff in a one-shot classical Minority
game, where the the players can do no better than choosing
betweeen the two options with equal probability.
This is as expected since in the absence
of entanglement the QMG cannot give any advantage over its
classical counter part.

We can search for a NE in the case of general $\alpha$ as follows.
If there exists $\theta^{*}, \, \beta^{*}$ such that
\begin{equation}
\label{eqn:symNE}
\left\langle \$_\text{D} \left( \hat{M}(\theta^{*}, \beta^{*}, -\beta^{*})^{\otimes 4} \right) \right\rangle \ge
 \left\langle \$_\text{D} \left( \hat{M}(\theta^{*}, \beta^{*}, -\beta^{*})^{\otimes 3}
    \otimes \hat{M}(\theta, \beta, -\beta) \right) \right\rangle \quad \forall\; \theta, \beta,
\end{equation}
where $\$_\text{D}( \cdot )$ denotes the payoff to the fourth player (Debra)
for the indicated strategy profile,
then $\hat{M}(\theta^{*}, \beta^{*}, -\beta^{*})$ is a symmetric
NE strategy. There is no in principle objections to asymmetric NE
strategy profiles, where the players choose different strategies,
however in practice these cannot be reliably achieved in the
absence of communication between the players since it is otherwise
impossible for the players to know which of the different
strategies to select. Necessary, but not sufficient, conditions
for the existence of a symmetric NE are
\begin{subequations}
\begin{align}
\label{eqn:deriv} \left. \frac{d \langle \$_\text{D} \rangle}{d
\theta} \right|_{\theta = \theta^{*}, \: \beta = \beta^{*}} &=0,
 & \left. \frac{d \langle \$_\text{D} \rangle}{d \beta} \right|_{\theta = \theta^{*}, \: \beta = \beta^{*}} &= 0, \\
\label{eqn:deriv2} \left. \frac{d^2 \langle \$_\text{D} \rangle}{d
\theta^2} \right|_{\theta = \theta^{*}, \: \beta = \beta^{*}} &\le
0,
 & \left. \frac{d^2 \langle \$_\text{D} \rangle}{d \beta^2} \right|_{\theta = \theta^{*}, \:  \beta = \beta^{*}} &\le 0,
\end{align}
\end{subequations}
where $\langle \$_\text{D} \rangle$ is here the payoff on the right
hand side of \eqref{eqn:symNE}. Inequalities in \eqref{eqn:deriv2}
indicate a local maximum in the payoff to Debra, however this may
not be a global maximum. An equality in \eqref{eqn:deriv2} may
mean a local maximum, minimum or an inflection point in the
payoff. We shall now enumerate all the symmetric NE strategies by
considering Eqs.~(\ref{eqn:deriv}--\ref{eqn:deriv2}) over the
range of $\alpha \in [0,1]$.

\begin{figure}
\includegraphics[scale=0.8]{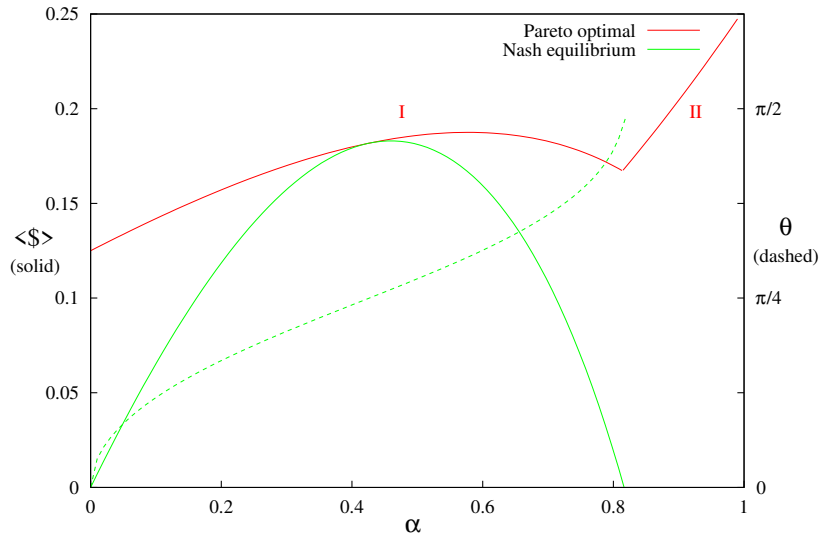}
\caption{\emph{Left scale, solid lines}: The Pareto optimal (\textcolor[rgb]{1,0,0}{-$\bullet$-})
and Nash equilibrium (\textcolor[rgb]{0,1,0}{-$\bullet$-}) payoffs as a function of the initial state parameter $\alpha$.
The Pareto optimal payoff curve is labeled by I or II for strategy $\hat{M}(\frac{\pi}{4}, 0, 0)$
or $\hat{M}(\frac{\pi}{2}, \frac{\pi}{8}, -\frac{\pi}{8})$, respectively.
\emph{Right scale, dashed line}: The value of $\theta$ given by \eqref{eqn:ttilde} that,
along with $\beta=0$, gives a symmetric Nash equilibrium.}
\label{fig:extremum}
\end{figure}

If Alice, Bob and Charles use the strategy $\hat{M}(\theta, \beta,
-\beta)$ while Debra plays $\hat{M}(\theta', \beta', -\beta')$
then
\begin{equation}
\begin{split}
\left. \frac{d \langle \$_{\rm D} \rangle}{d \theta'}
\right|_{\theta' = \theta, \: \beta' = \beta}
 &= \frac{\sin 2 \theta}{8} \left[ 2 \alpha^2 \,+\, 2 \alpha \sqrt{2 - 2 \alpha^2} \, \cos 4 \beta \right. \\
 &\left. \quad +\, (2 \alpha^2 - 2 - 2 \alpha \sqrt{2 - 2 \alpha^2} \, \cos 4 \beta
 - \alpha^2 \cos^2 4 \beta) \sin^2 \theta \right], \\
\left. \frac{d \langle \$_{\rm D} \rangle}{d \beta'}
\right|_{\theta' = \theta, \: \beta' = \beta}
 &= \frac{\alpha}{2} \sin 4 \beta \, \sin^2 \theta \\
 &\quad \times \left[ (\sqrt{2 - 2 \alpha^2} + \alpha \cos 4 \beta) \sin^2 \theta
 - 2 \sqrt{2 - 2 \alpha^2} \right].
\end{split}
\end{equation}
We want to find $\theta, \beta$ for which these derivatives are
simultaneously zero. Apart from the known NE for $\alpha=1$ we
find that the only other symmetric NE occurs for $\alpha \le
\sqrt{\frac{2}{3}}$ when $\cos 4 \beta = 1$  and
\begin{equation}
\label{eqn:ttilde} \cos {\theta} =
    \sqrt{\frac{2 - 3 \alpha^2}{2 - \alpha^2 + 2 \alpha \sqrt{2 - 2 \alpha^2}}}.
\end{equation}
The expected payoff to each player for this equilibrium is
\begin{equation}
\label{eqn:NEpay4} \langle \$ \rangle = \frac{\alpha (2 - 3
\alpha^2) (\alpha + \sqrt{2 - 2 \alpha^2})}
                                  {4 - 2 \alpha^2 + 4 \alpha \sqrt{2 - 2 \alpha^2}},
\end{equation}
which reaches a maximum value of $(3 + 2 \sqrt{3})/(18 + 10
\sqrt{3}) \approx 0.183$ at $\alpha = \sqrt{\frac{1}{6}(3 -
\sqrt{3})}$.
Figure~\ref{fig:extremum} gives
the value of $\theta$ and the resulting payoff for this solution.

We now consider the PO strategy profile.
Again we will only consider symmetric strategy
profiles. That is, we are interested in $\theta^{*}, \, \beta^{*}$
for which
\begin{equation}
\label{eqn:symPO}
\left\langle \$ \left( \hat{M}(\theta^{*},
\beta^{*}, -\beta^{*})^{\otimes 4} \right) \right\rangle \ge
\left\langle \$ \left( \hat{M}(\theta, \beta, -\beta)^{\otimes 4}
\right) \right\rangle \quad \forall\; \theta, \beta,
\end{equation}
where $ \$ $ represents the payoff to any one of the four players
for the indicated strategy profile. Suppose all players select the
strategy $\hat{M}(\theta, \beta, -\beta)$ for some $\theta, \,
\beta$ to be determined.
We proceed as before by finding the stationary points of the payoff
to each player as a function of the parameters $\theta$ and $\beta$.
We find that for $\alpha > \sqrt{\frac{2}{3}}$,
where the initial state is dominated by the GHZ state,
the optimal strategy is $\hat{M}(\frac{\pi}{2}, \frac{\pi}{8}, -\frac{\pi}{8})$,
while for the EPR-dominated region, $\alpha < \sqrt{\frac{2}{3}}$,
the optimal strategy is $\hat{M}(\frac{\pi}{4}, 0, 0)$.
At $\alpha = \sqrt{\frac{2}{3}}$ all components in the
initial state are equally weighted and both strategies yield the
same results.
The payoffs for the two regions are
\begin{align}
\langle \$ \rangle_{\rm I} &= \frac{1}{8} + \frac{f}{16} \, \alpha (2 \sqrt{2 - 2 \alpha^2} - \alpha)
    & \alpha \le \sqrt{\frac{2}{3}} \\
\langle \$ \rangle_{\rm II} &= \frac{1}{8} + \frac{f}{8} \, \alpha (2 \alpha^2 - 1)
    & \alpha \ge \sqrt{\frac{2}{3}}
\end{align}
Figure~\ref{fig:extremum} shows the payoffs for these
cases for a fidelity of $f=1$.

There has been some recent interest in the correspondence between
equilibria in quantum game theory and the violation of Bell
inequalities~\cite{Sil07,Che07}. In our case we note that the
curve for the symmetric PO payoff is the same as that for the
maximal violation of the Mermin-Ardehali-Belinski-Klyshko (MABK)
inequality~\cite{Fli08}.

\section{Experiment}
\label{sec:experiment}

\subsection{State observation:}\label{sec:setup}
The states are obtained with a new linear optics network which
enables the observation of a whole family of states in a single
setup by the tuning of one experimental parameter \cite{Wie08}.
The setup relies on the interference of the second order emission
of type II non-collinear spontaneous parametric down conversion
(SPDC) and is depicted in Fig.~\ref{fig:setup}. The down
conversion emission yields four photons, two horizontally and two
vertically polarized, in two spatial modes $a$ and $c$. The
photons are overlapped on a polarizing beam splitter (PBS) and
afterwards symmetrically split up by two polarization independent
beam splitters (BS) into the four spatial modes
$\{a,\,b,\,c,\,d\}$. Prior to the interference at the PBS the
polarization of the photons in mode $c$ is rotated by a
half-waveplate (HWP). Under the condition of detecting one photon
in each spatial mode the desired state is observed. The weighting
coefficient $\alpha$ is thereby determined by the orientation of
the principal axis $\gamma$ of the HWP according to
\begin{equation}\label{eqn:alpha2angle}
        \alpha(\gamma)=\frac{2\sqrt{2}\sin^2(2\,\gamma)}{\sqrt{5-4\cos(4\,\gamma)+3\cos(8\,\gamma)}}
        \end{equation}
with $\gamma\in[0,\frac{\pi}{8}]$.

\subsection{Implementation of the game:}
The implementation of the four-player QMG
consists of acting with the strategy operator on each qubit and
afterwards measuring the resulting output state in different
bases. The unitary transformation corresponding to the players'
choice of strategy is realized by a series of quarter-, half-
and quarter-waveplate (see Fig.~\ref{fig:setup}). In order to play
strategy I and to act with $\hat M_{\textsc{i}}\equiv\hat
M(\frac{\pi}{2},\frac{\pi}{8}, -\frac{\pi}{8})$ the angles of the waveplates are
chosen as $\{-\frac{\pi}{8},\, \frac{5\pi}{16},\,0\}$, and for
strategy II and $\hat M_{\textsc{ii}}\equiv\hat
M(\frac{\pi}{4},0,0)$ as $\{\frac{\pi}{2},\,
\frac{\pi}{16},\,\frac{\pi}{2}\}$.

\subsection{Results:}
We start the measurement with $\alpha=1$ as in this case a NE
solution is expected. The expected
state after the application of strategy I, which is here the
optimal one, is of the form
\begin{eqnarray}
(\hat M_{\textsc{i}}^{\otimes
4})\ket{GHZ}&=&-\frac{1}{2\sqrt{2}}\Bigl(\ket{HHHV}+\ket{HHVH}
+\ket{HVHH}+\ket{VHHH}\notag\\
& &-\ket{VVVH}-\ket{VVHV}
-\ket{VHVV}-\ket{HVVV}\Bigr)\notag\\
&=&\frac{i}{\sqrt{2}}(\ket{RRRR}-\ket{LLLL}),
\end{eqnarray}
with $\ket{R/L}=\frac{1}{\sqrt{2}}(\ket{H}\pm i\ket{V})$
being the right and left circular polarization states.
The transformed state is also of GHZ-type, and thus the fidelity
$F=\bra{GHZ}M_{\textsc{i}}^{\dagger\otimes
4}\rho^{\alpha=1}_{\mathrm{exp}}\hat M_{\textsc{i}}^{\otimes
4}\ket{GHZ}$ of the experimental state
$\rho^{\alpha=1}_{\mathrm{exp}}$ equals the average expectation
value of the GHZ stabilizer operators~\cite{Kie05}. A measurement
of the respective correlations is hence sufficient to evaluate the
state fidelity. In our case, this requires 9 measurement settings
and we obtain a value of $F=0.746 \pm 0.019$. For comparison, the
fidelity of the untransformed initial GHZ state obtained with our
setup is $F=0.745\pm0.022$.

The payoff awarded to each player can be evaluated from the
correlation measurement in the computational basis,
$\sz\otimes\sz\otimes\sz\otimes\sz$.
Averaged over all four players the
payoff obtained for $\alpha=1$ is $\langle \$_{\textsc{i}}
\rangle_{\alpha=1}=0.206\pm 0.009$ which is well above the
classical limit of $\frac{1}{8}$.

\begin{figure}
\includegraphics[width=8.5cm,keepaspectratio]{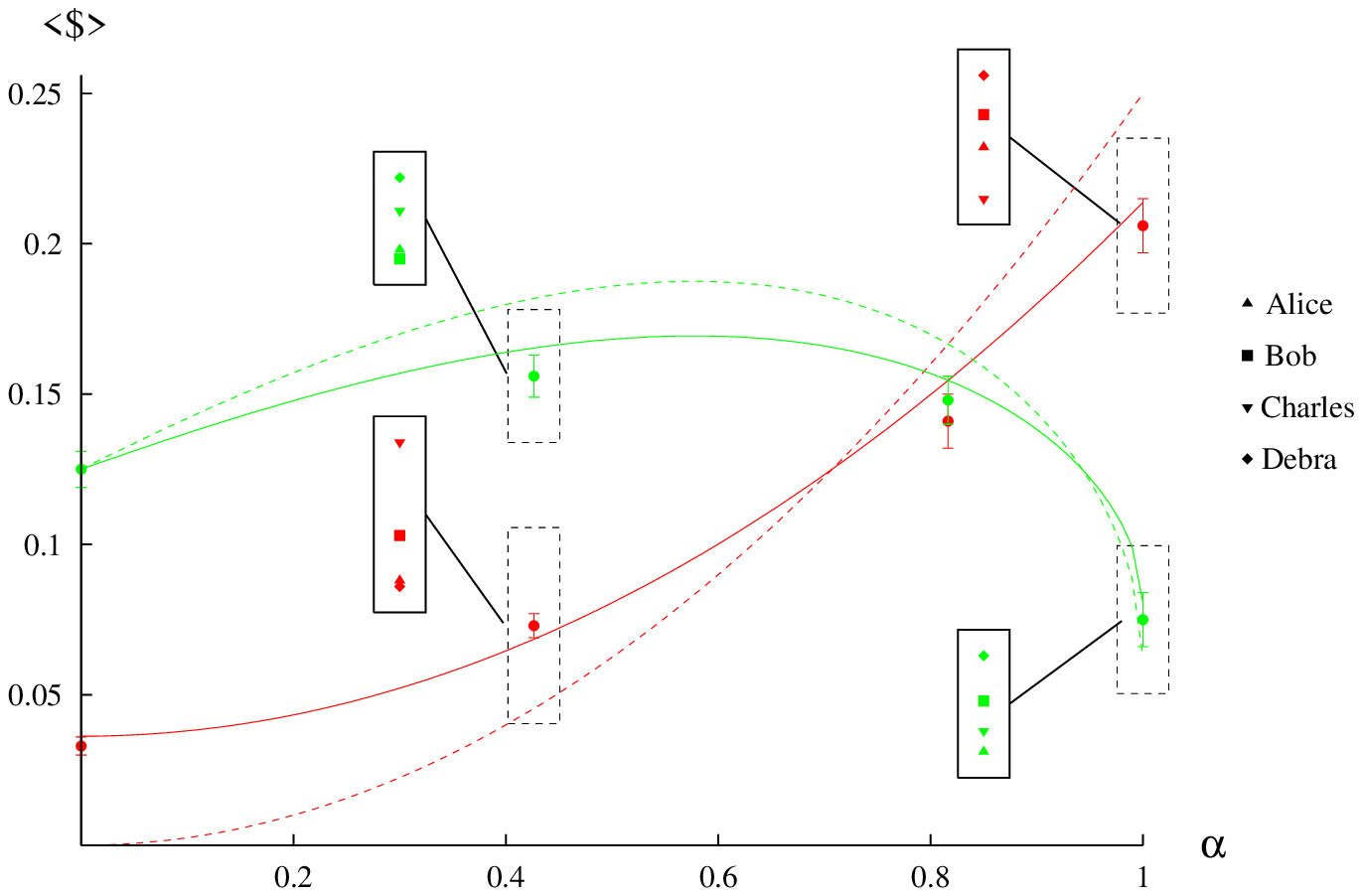}
\caption{Measured payoff $\langle \$_{\textsc{i},\textsc{ii}}
\rangle$ averaged over all four players as a function of $\alpha$ for
strategy I (\textcolor[rgb]{1,0,0}{-$\bullet$-}) and strategy II
(\textcolor[rgb]{0,1,0}{-$\bullet$-}). The data is fitted assuming
a mixed input state of the form
\mbox{$f\,\ketbra{\Psi(\alpha)}+(1-f)/16 \,\openone^{\otimes 4}$}
[see \eqref{eqn:noise}].
Dashed lines correspond to the ideal case $f=1$.
The boxes show the individual values for the four players for some sample points.
Typical errors for individual measurements are 9--12 \%.
The maximum equilibrium payoff in a classical game is $\frac{1}{8}$.}
\label{fig:povsalpha}
\end{figure}

As was shown earlier, the maximal achievable
payoff depends on $\alpha$ and, moreover, there is an interesting
change in the Pareto optimal strategy at
$\alpha=\sqrt{2/3}$. In order to test this experimentally
we chose to perform measurements 
for other distinguished states, $\ket{\Psi(0)}$,
$\ket{\Psi(\sqrt{\frac{2}{11}})}$, and
$\ket{\Psi(\sqrt{\frac{2}{3}})}$.
The results obtained are
summarized in Fig.~\ref{fig:povsalpha}
which gives the average PO payoffs for the four values of $\alpha$ for strategies I and II,
along with best fit curves and theoretical curves for perfect fidelity.
As can be seen in the boxes at some example points, the payoffs
differ slightly for each player but are generally comparable within typical
measurement error of around 9--12\%.
The only exception occurs for player $C$ and player $A$ for
$\alpha=\sqrt{2/11}$ and $\sqrt{2/3}$, respectively.
The cause of this could not be rigorously identified. The average payoffs
follow the expected dependence for both strategies, once imperfect
state quality is taken into account. To allow for loss in the
states' fidelity the data in Fig.~\ref{fig:povsalpha} is fitted
assuming a mixed input state of the form \eqref{eqn:noise}.
Although the assumptions of an admixture of white noise and
constant state quality are only an approximations, the value of
$f=0.71 \pm 0.03$ obtained from the fit is in good agreement with
the measured GHZ fidelity, where $F=(1+15f)/16\approx0.73$. Let us
discuss the results for each measured point in more detail.

For $\alpha=0$ a product of two Bell states is expected. As this
state is not four-qubit entangled it should yield at best the
classical payoff for strategy II. This is nicely reproduced in the
experiment with an average payoff of $\langle \$_{\textsc{ii}}
\rangle_{\alpha=0}=0.125\pm0.006$. If the players chose to play
strategy I zero payoff is expected for ideal, pure input states.
However, for an increasing mixedness of the input states the payoff
approaches the classical limit, with equality for $f\rightarrow 0$.
This fact leads to a non-zero value for the experimentally measured
average payoff for strategy I, $\langle \$_{\textsc{i}}
\rangle_{\alpha=0}=0.033\pm0.003$.

Similar behavior is obtained for $\alpha=\sqrt{2/11}$, where the
measured payoffs for strategy II, $\langle \$_{\textsc{ii}}
\rangle_{\alpha=\sqrt{2/11}}=0.156\pm0.007$, and strategy I,
$\langle \$_{\textsc{i}}
\rangle_{\alpha=\sqrt{2/11}}=0.073\pm0.004$, are lower and higher
than expected, respectively. Never the less, the experimentally reached
state quality is high enough to ensure for the proper strategy, a
payoff which exceeds the one maximally achievable in the equivalent classical game.

As mentioned before, $\alpha=\sqrt{2/3}$ is
special as both strategies lead to the same payoff. It is the
point about which the players have to switch between the different
strategies. The corresponding state,
$\ket{\Psi(\sqrt{2/3})}\equiv\ket{\Psi_4}$ is well known from
literature as it can be directly obtained from SPDC
\cite{Wei01,Eib03,Gae03} and has several interesting applications
in quantum information (see, for example, Refs.~\cite{Bou04,Mur99}).
The point's feature of being a quantum fulcrum is nicely reproduced in the
experiment. Within the measurement errors the average payoffs, $\langle \$_{\textsc{i}}
\rangle_{\alpha=\sqrt{2/3}}=0.141\pm0.009$ and $\langle
\$_{\textsc{ii}} \rangle_{\alpha=\sqrt{2/3}}=0.148\pm0.008$ are the
same for strategy I and II
(see Fig.~\ref{fig:povsalpha}).

Finally, the values obtained in the GHZ case fit well in the
dependence prescribed by the previous points. Like for the other
states, due to imperfect state quality the average payoff is
slightly lower (higher) than the ideal value of $\frac{1}{4}$ ($\frac{1}{16}$) for
strategy I (II).

In summary, we can state that for all values of $\alpha$ the
players are awarded payoffs above the classical NE value if they
choose the appropriate strategy.
The values are consistent with a fidelity of 0.73
for the initial production of four-partite entangled state.

In our realization of the Minority game we do not use an
unentangling gate. Thus a measurement in the computational basis
alone does not prove that the higher than classical payoff values
have their seeds in the four-qubit entanglement of
$\ket{\Psi(\alpha)}$. In principle they could be caused by an
admixture of the separable state
\begin{eqnarray}
\rho_{\textrm{sep}}&=&\frac{1}{8}\Big(\ketbra{HHHV} +\, \text{permutations}\notag \\
&& + \ketbra{VVVH}+\,\text{permutations}\Big),
\end{eqnarray}
which yields a maximal ``payoff'' of $\frac{1}{4}$ when measured in the
computational basis. However, measured in other bases, the state
$\rho_{\mathrm{sep}}$ will not give a payoff above the classical
limit.

\begin{figure}
\includegraphics[width=8.5cm,keepaspectratio]{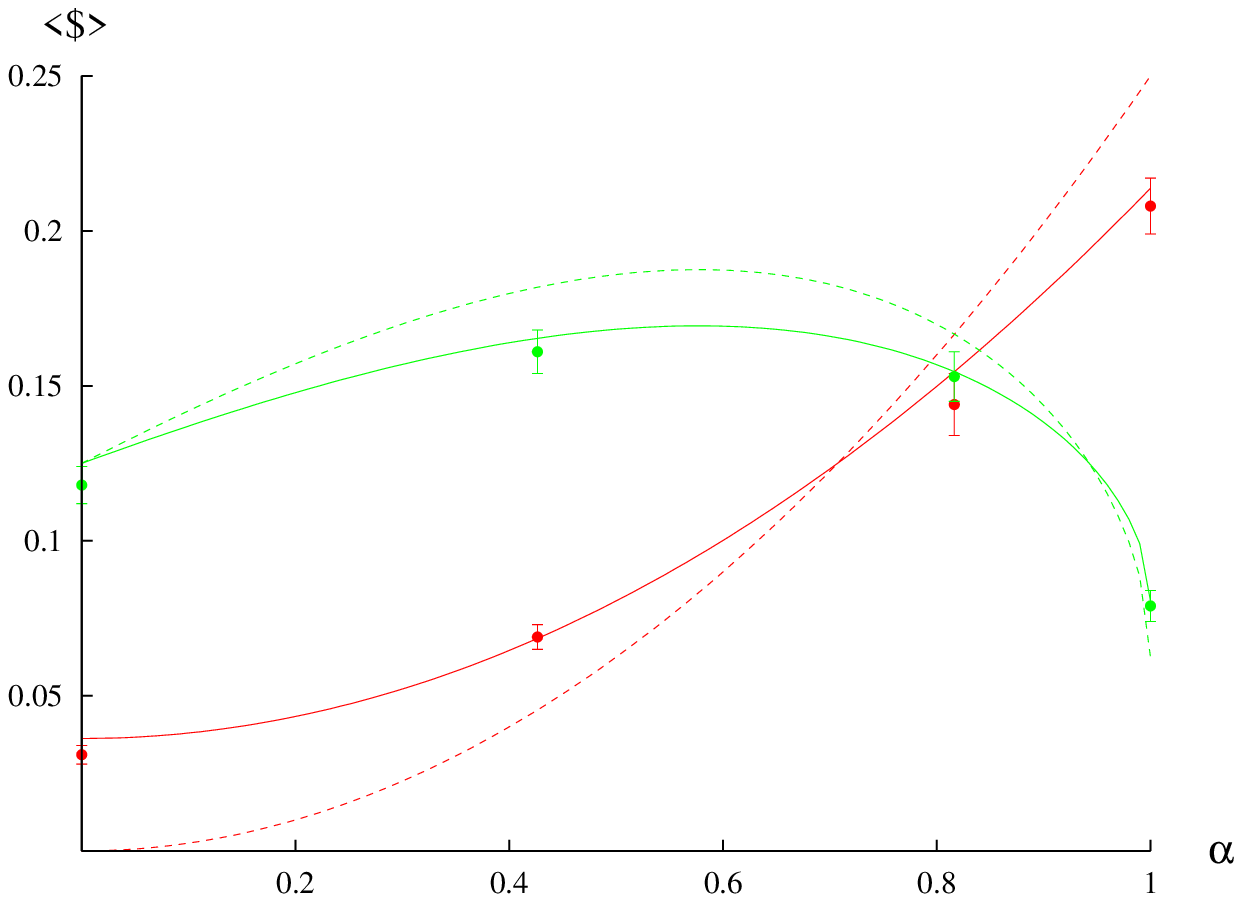}
\caption{Measured payoff $\langle \$_{\textsc{i},\textsc{ii}}
\rangle$ averaged over all four players as a function of $\alpha$ in
the basis $\sx \otimes \sx \otimes \sx \otimes \sx$ for strategy I
(\textcolor[rgb]{1,0,0}{-$\bullet$-}) and in the basis
$\sy \otimes \sy \otimes \sy \otimes \sy$ for strategy II
(\textcolor[rgb]{0,1,0}{-$\bullet$-})
The data is fitted assuming a mixed input state of the form
\mbox{$f\,\ketbra{\Psi(\alpha)}+(1-f)/16 \,\openone^{\otimes 4}$}
[see \eqref{eqn:noise}].
Dashed lines correspond to the ideal case $f=1$.
The maximum average equillibrium payoff in a classical game is $\frac{1}{8}$.}
\label{fig:povsalphaxy}
\end{figure}

In contrast, the state $\ket{\Psi(\alpha)}$ has the extraordinary
property that it exhibits the same term structure in the bases
$\sz \otimes \sz \otimes \sz \otimes \sz$ and $\sx \otimes \sx
\otimes \sx \otimes \sx$ when transformed by $\hat M^{\otimes
4}_{\textsc{i}}$ and in the bases $\sz \otimes \sz \otimes \sz
\otimes \sz$ and $\sy \otimes \sy \otimes \sy \otimes \sy$ when
transformed by $\hat M^{\otimes 4}_{\textsc{ii}}$. Consequently,
if the payoff is evaluated analogously in these bases, the same
dependence on $\alpha$ is expected. In order to prove this we have
performed the same measurements as before in the bases $\sx
\otimes \sx \otimes \sx \otimes \sx$ and $\sy \otimes \sy \otimes
\sy \otimes \sy$. The result is shown in
Fig.~\ref{fig:povsalphaxy}.

For each basis and appropriate strategy we find
very similar curves to those for the computational basis.
We applied the same fitting procedure as before. The resulting values for $f$
are slightly different for both bases ($f_x=0.736\pm 0.019$ and
$f_y=0.706\pm0.053$) but comparable with the one found for the
computational basis. The small asymmetry between the bases indicates that,
contrary to our assumption,
the experimental noise is not purely white noise.

\section{Conclusions}
We have studied the four-player quantum Minority game
where the initial state is drawn from a continuous set of four-partite entangled states
consisting of a superposition of the GHZ state and a product of EPR pairs.
There is a well known Nash equilibrium when the initial state is a GHZ state.
For other initial states in our set there are only symmetric Nash equilibria when the initial state
is in the EPR-dominated region,
though the payoff for this equilibrium is only sometimes higher than that achievable classically.
There are two symmetric Pareto optimal strategies,
one for the EPR- and one for the GHZ-dominated initial state.
In both cases the payoff is equal to the classical optimal payoff of $\frac{1}{8}$
plus a value proportional to the fidelity to which the initial entanglement is produced.

An experimental implementation of this game is carried out using four-photon entangled states
generated from spontaneous parametric down conversion.
Results for four different initial states were obtained.
The results are consistent with the theory for fidelities in the range 0.71--0.73 depending on the initial state.
For these fidelities the average payoffs obtained are above those that can be obtained
in a classical Minority game (with an unentangled initial state)
thus demonstrating the usefulness of entanglement.
This is the first implementation of a multiplayer quantum game.

    The study of quantum games is interesting in general as it
    provides another point of view to exploit pre-existing classical
    frameworks as a base for finding new ways of understanding and
    using entanglement in quantum systems. While game theory is
    the  mathematical language of competitive (classical)
    interactions, quantum game theory is the natural extension to
    consider competitive situations in quantum information
    settings. For example, eavesdropping in quantum communication
    (see, for example, Refs.~\cite{Gis97a, Eke91, Ben92a})
    and optimal cloning \cite{Wer98} can be conceived as strategic games between
    two or more players. The particular game discussed here represents a different means of studying
    the entanglement structure of the used input states. As shown
    elsewhere \cite{Fli08}, there even exists an interesting correspondence
    between the equilibria in the Minority game and the violation
    of Bell inequalities.

\section{Methods}

For operation of the SPDC a $\beta$-Barium borate (BBO) crystal is
pumped by UV pulses having a central wavelength of 390~nm and an
average power of 600~mW obtained from a frequency-doubled
Ti:Sapphire oscillator (pulse length is 130~fs). Another HWP and a
1~mm thick BBO crystal compensate walk-off effects. The spatial
modes $a$ and $c$ are defined by coupling the emitted photons in
single mode fibres. Spectral filtering of each mode is achieved by
interference filters centered around the degenerate wavelength of
780~nm with 3~nm full width at half maximum. The optical path
lengths in modes $a$ and $c$ are aligned by a Hong--Ou--Mandel (HOM)
interference measurement \cite{Hon87} to ensure temporal mode
matching. Pinholes serve as mode filters to enhance the spatial
indistinguishability of the photons at the PBS. Phases between
horizontal and vertical polarization are set by pairs of
perpendicularly oriented birefringent Yttrium-Vanadate (YVO$_4$)
crystals. Polarization analysis is performed in all of the four
outputs by half- and quarter-waveplates followed by a PBS. The
photons are detected by eight Silicon avalanche photo diodes
(Si-APD) whose signals are fed into a multichannel coincidence unit
which allows us to simultaneously register any possible coincidence
detection between the inputs. The rates for each of the 16
characteristic fourfold coincidences are corrected for the
difference in the efficiencies of the detectors. The errors on all
quantities are deduced from Poissonian counting statistics of the
raw detection events and independently determined efficiencies.

\bibliography{bibliography}


\section{Acknowledgements}
This work was supported by the Deutsche Forschungsgemeinschaft and
the European Commission through the EU Project QAP.
APF was funded by an Australian Research Council (ARC) Postdoctoral
Fellowship project number DP0559273 and LCLH is the recipient of an
ARC Professorial Fellowship project number DP0770715.

[Competing Interests] The authors declare that they have no
competing financial interests.

[Correspondence] Correspondence and requests for materials should be
addressed to CS~(email: christian.schmid@mpq.mpg.de).

\end{document}